\documentclass[paper=a4, fontsize=11pt]{scrartcl} 
\pdfoutput=1

\usepackage[T1]{fontenc} 
\usepackage{color}
\usepackage[english]{babel} 
\usepackage{amsmath,amsfonts,amsthm} 
\usepackage{graphicx}
\usepackage{hyperref}
\allowdisplaybreaks

\numberwithin{equation}{section} 
\numberwithin{figure}{section}
\numberwithin{table}{section} 
\begin{document}

\begin{titlepage}
\begin{center}

{\Large {\textbf{
$\beta$-function formalism for inflationary models with a non minimal coupling with gravity.}}}
 
 ~\\
\vskip 2cm

\vskip 1cm

{  \textbf{M. Pieroni$^{ab}$ } }\\
~\\
~\\
{\em ${}^a$ Laboratoire AstroParticule et Cosmologie,
Universit\'e Paris Diderot } \\
{\em ${}^b$ Paris Centre for Cosmological Physics, F75205 Paris Cedex 13}\\

\end{center}

\vskip 1cm
\centerline{ {\textbf{ Abstract}}}{ We discuss the introduction of a non minimal coupling between the inflaton and gravity in terms of our recently proposed $\beta$-function formalism for inflation. Via a field redefinition we reduce to the case of minimally coupled theories. The universal attractor at strong coupling has a simple explanation in terms of the new field. Generalizations are discussed and the possibility of evading the universal attractor is shown. }

\indent

\vfill

\vfill

\end{titlepage}
\date{\normalsize\today} 

\tableofcontents
\section{Introduction}
Inflation is the most suitable extension of standard cosmology to solve the horizon, monopoles and flatness problems. The Planck mission \cite{Ade:2015lrj} and other cosmological observations help to fix several constraints on the general mechanism driving this phenomenon. The chaotic model \cite{Linde:1983gd} with potential $V(\phi) = \lambda \phi^4$ with a non-minimal coupling of the scalar field with gravity $\frac{\xi \phi^2}{2} R$ has been recently proposed by Bezrukov and Shaposhnikov \cite{Bezrukov:2007ep} as a natural extension of the Standard Model in order to include inflation. For a large number $N$ of e-folding this model gives predictions for the scalar spectral index and the tensor to scalar ratio:
\begin{equation}
\label{ns-r-attractor}
n_s = 1 - \frac{2}{N}, \qquad \qquad \qquad r = \frac{12}{N^2}.
\end{equation}
Assuming that $N\sim50-60$ we find numerical values in good agreement with Planck data. As Starobinsky model \cite{Starobinsky:1980te} and many other inflationary models are also predicting similar values for $n_s$ and $r$ it is important to define a systematic classification in order to avoid this degeneracy. Some proposals to explain this degeneracy have been formulated by Mukhanov \cite{Mukhanov:2013tua} and Roest \cite{Roest:2013fha}. In this spirit we recently proposed a $\beta-$function formalism for inflation \cite{Binetruy:2014zya}. This new approach is based on the idea of providing universality classes of models of inflation by relying on the approximate scale invariance during the inflationary epoch. This suggestion has a deep connection with the idea proposed by McFadden and Skenderis \cite{McFadden:2010na} of applying the holographic principle to describe the inflationary Universe. In the language of the well known (A)dS/CFT correspondence of Maldacena \cite{Maldacena:1997re}, the asymptotic de Sitter spacetime is dual to a (pseudo) Conformal Field Theory. In this framework the equations describing the cosmological evolution are thus interpreted as holographic Renormalization Group (RG) equations for the corresponding QFT \cite{Kiritsis:2013gia}.
This correspondence suggests that universality classes for inflationary models should be defined in terms of the Wilsonian picture of fixed points (exact deSitter solutions), scaling regions (inflationary epochs), and critical exponents (scaling exponents of the power spectra related with the slow roll parameters). It is important to stress that, in analogy with statistical mechanics, these universality classes should be considered as sets of theories that share a common scale invariant limit. As results obtained in this framework are not only valid for particular models but for whole sets of theories, it should be clear that they should be conceived as more general than the ones obtained using the standard methods. \\

 In this paper we discuss inflationary models where a scalar field is non-minimally coupled with gravity. A discussion of this topic has been recently proposed by Linde, Kallosh and Roest \cite{Kallosh:2013tua} in terms of the standard picture of defining inflationary models by identifying the inflationary potentials and they proved the existence of a universal attractor at strong coupling. Both to have a deeper comprehension of the inflationary regime and to produce a further generalization of the results presented in \cite{Kallosh:2013tua}, it is interesting to treat theories for scalar field with a non-minimal coupling with gravity in terms of the $\beta-$function formalism. In Sec.\ref{sec:model_definition}, we present a model of a scalar field with a non-minimal coupling with gravity. In Sec.\ref{sec:beta_function} we formulate the problem in terms of the $\beta-$function formalism and we present the weak and strong limits. In Sec.\ref{sec:general_case}, we consider a more general class of models by relaxing an assumption on the expression for the potential. In this context we prove that it is possible to evade the universal attractor and that other attractors can be reached. In Sec.\ref{sec:conclusions}, we finally present our conclusions.

\section{Setting up the model}
\label{sec:model_definition}
The simplest action to describe the inflating universe consists of a the standard Einstein-Hilbert term to describe gravity plus the action for a homogeneous scalar field in curved spacetime\footnote{We use the convention $ds^2 = dt^2 - a(t)^2 (dr^2 + r^2 d\Omega^2)$}:
  \begin{equation}
  \label{minimal-action}
    S =\int\mathrm{d}^4x\sqrt{-g}\left( -  \frac{1}{2\kappa^2}R +  X - V_J(\phi) \right),
  \end{equation}
where $X\equiv g^{\mu \nu}  \partial_\mu \phi \partial_\nu \phi /2 = \dot{\phi}^2 /2 $ is the standard kinetic term for a homogeneous scalar field. Let us consider a generalization of this action to include a non-minimal coupling between the scalar field and gravity. In this paper, we follow the proposal of \cite{Kallosh:2013tua}, and we consider the action:
    \begin{equation}
  \label{non-minimal-action2}
    S =\int\mathrm{d}^4x\sqrt{-g}\left( -  \frac{\Omega(\phi)}{2\kappa^2}R +  X - V_J(\phi) \right).
  \end{equation}
As gravity is not described by a standard Einstein-Hilbert term, this should be considered as the Jordan frame formulation of the model. Notice that we have not imposed any constraint on the explicit expression of $V_J(\phi)$. Let us consider: 
  \begin{equation}
\label{omega}
  \Omega(\phi) = 1 + \xi f(\phi),
  \end{equation}
where $\xi$ is the coupling constant and $f(\phi)$ is a function of $\phi$. It is again interesting to stress that this parametrization is quite general as we are not imposing any constraint on the explicit expression for $f(\phi)$. It should also be stressed that $\xi = 0$ corresponds to the standard case of a scalar field minimally coupled with gravity.\\
It is well known that by means of a conformal transformation i.e.
  \begin{equation}
\label{metric}
  g_{\mu\nu} \rightarrow \Omega(\phi)^{ -1} g_{\mu\nu},
  \end{equation}
we can recover the standard Einstein-Hilbert term for gravity. The action in terms of the new metric can be expressed as:
    \begin{equation}
  \label{non-minimal-action}
    S=\int\mathrm{d}^4x\sqrt{-g}\left( - \frac{1}{2\kappa^2}R +  F(\phi)X - \bar{V}(\phi) \right),
  \end{equation}
where $F(\phi)$ and $\bar{V}(\phi)$ are defined by:
  \begin{equation}
\label{generalF}
    F(\phi) \equiv \Omega^{-1} + \frac{3}{2} \left( \frac{\mathrm{d} \ln \Omega}{\mathrm{d} \phi}\right)^2, \qquad \qquad \qquad \bar{V}(\phi) \equiv \frac{V_J (\phi)}{\Omega(\phi)^2}. 
  \end{equation}
This is usually known as the Einstein frame formulation for the theory. From now on, we impose $\kappa^2 = 1 $ to simplify the notation. As discussed in \cite{Kallosh:2013tua}, it is interesting to consider the particular expression for $V_J(\phi)$:
\begin{equation}
V_J(\phi) = \lambda^2 f^2(\phi).
\end{equation}
This parametrization is motivated by the possibility of defining a natural supergravity embedding \cite{Kallosh:2010ug} for this class of models. It is important to stress that in the limit of small coupling $\xi \ll 1 $, both $\Omega(\phi)$ and $F(\phi)$ are close to one. In this limit, fixing an explicit parametrization for $f(\phi)$ we are directly fixing the potential for the theory. At this point it should be clear that in this regime different choices for $f(\phi)$ correspond to different predictions for $n_s$, scalar spectral index, and $r$, tensor to scalar ratio. As discussed in \cite{Kallosh:2013tua}, it is interesting to consider the limit of a strong coupling $1 \ll \xi$. It possible to show that in this regime the expression for $N$, number of e-foldings, simply reads:
  \begin{equation}
\label{e-foldings}
    N(\phi) \simeq \frac{3}{4} \xi f(\phi). 
  \end{equation}
It is also possible to show that in this limit, the expressions for $n_s$ and $r$ are simply given by eq.\eqref{ns-r-attractor}. It is important to stress that this result is independent on the explicit choice for $f(\phi)$. As different theories share the same asymptotic behavior in the limit of $1 \ll \xi$, this proves the existence of a universal attractor at strong coupling. In the rest of this work we will focus both on the interpretation of this attractor in terms of the $\beta$ function formalism of \cite{Binetruy:2014zya}, and on the possibility of extending these results for more general classes of models. In particular, in Sec.\ref{sec:general_case}, we will discuss the consequences of chosing a different parametrizetion for $V_J(\phi)$ i.e. 
\begin{equation}
\Omega(\phi) = 1 + \xi f(\phi) ,\qquad \qquad \qquad V_J(\phi) = \lambda^2 g^2(\phi) ,
\end{equation}
with $f(\phi) \neq g(\phi)$.

\section{$\beta$-function formalism}
\label{sec:beta_function}
Let us consider the model described in Sec.\ref{sec:model_definition}. By means of a field redefinition it is possible to reduce to the problem of a scalar field with a canonically normalized kinetic terms. In particular let us define\footnote{ Notice that this definition may drive to an ambiguity as it implies $d\varphi / d \phi = \pm \sqrt{F(\phi)}$. To solve this problem we simply have to select a solution and be consistent with our choice. In this paper we will consider the $+$ solution. Clearly an equivalent treatment can be achieved in terms of the $-$ solution. } the new field $\varphi$ as:
\begin{equation}
\label{def_varphi}
\left( \frac{\mathrm{d} \varphi}{ \mathrm{d} \phi} \right)^2 = F(\phi).
\end{equation} 
By definition the kinetic term of $\varphi$ is canonically normalized and thus we can directly follow the procedure discussed in \cite{Binetruy:2014zya}. Assuming that the time evolution of the scalar field $\varphi(t)$ is \emph{piecewise monotonic} we can invert to get $t(\varphi)$ and use the field as a clock. Under this assumption we can thus describe the dynamics of the system in terms of the Hamilton-Jacobi approach of Salopek and Bond \cite{Salopek:1990jq}. In this framework we define $W(\varphi) \equiv -2H(\varphi)$, that satisfies $\dot{\varphi} = W_{,\varphi} (\varphi)$ and also 
\begin{equation}
  \label{superpotential} 
  2 V(\varphi) = \frac{3}{2} W^2(\varphi) - \left[W_{,\varphi} (\varphi)\right]^2.
 \end{equation}
The latter expression leads to call the function $W(\varphi)$ \emph{superpotential} because of a similar parametrisation in the context of supersymmetry. In analogy with QFT we define:
\begin{equation}
  \label{beta_varphi} 
  \beta(\varphi) \equiv \frac{\mathrm{d} \varphi}{\mathrm{d} \ln a} = - 2\frac{\mathrm{d} \ln W(\varphi)}{\mathrm{d} \varphi}.
 \end{equation}
It is important to notice that the equation of state for the scalar field in terms of $\beta$ reads:
\begin{equation}
  \label{eq_of_state} 
 \frac{p+ \rho}{\rho} = \frac{\beta^2 (\varphi)}{3}.
 \end{equation}
This expression for the equation of state implies that an inflationary epoch is associated with the neighborhood of a zero of $\beta(\varphi)$. In fact, by specifying a parametrization for $\beta(\varphi)$, we are fixing the evolution of the system (or equivalently the RG flow) close to a fixed point. As a single asymptotic behavior can be reached by several models, the parametrization of $\beta(\varphi)$ is not simply specifying a single inflationary model but rather a set of theories sharing a scale invariant limit. In particular, using the language of statistical mechanics, we are specifying a \emph{universality class} for inflationary models. It is important to stress that in this framework all the informations on the inflationary phase are thus enclosed in the parametrization of $\beta(\varphi)$ in terms of the critical exponents. Substituting eq.\eqref{beta_varphi} into eq.\eqref{superpotential} we express the potential $V(\varphi)$ as:
  \begin{equation} 
\label{potential_1}
            V(\varphi)=\frac{3 W^{2}(\varphi)}{4}\left[1-\frac{\beta^{2}(\varphi)}{6}\right].
    \end{equation}
During an inflationary epoch $\beta(\varphi)$ must be close to zero and thus at the lowest order, we can approximate\footnote{As eq. \eqref{beta_varphi} implies: \begin{equation*}
W(\varphi) = W_\textrm{f} \exp \left[ - \int_{\varphi_f}^{\varphi} \frac{\beta(\hat{\varphi})}{2} d\hat{\varphi} \right], \end{equation*} to be consistent with this approximation we need: \begin{equation*} \left| \beta(\varphi)\right|^2 \ll \left| \int_{\varphi_f}^{\varphi} \beta(\hat{\varphi}) d\hat{\varphi} \right|. \end{equation*}
In the rest of this paper we will consider explicit expressions for $\beta(\varphi)$ that satisfy this requirement.} eq.\eqref{potential_1} with: $V(\varphi) \sim \frac{3}{4} W(\varphi)^2$. From eq. \eqref{eq_of_state}, we can notice that $\beta^2 /2 $ is equal to the first slow-roll parameter $\epsilon = - \dot{H}/H^2$. In the slow rolling regime the $\beta$-function formalism is thus equivalent to the horizon-flow approach of Hoffman and Turner \cite{Hoffman:2000ue, Kinney:2002qn, Liddle:2003py, Vennin:2014xta}. In this limit we can thus express $\beta(\varphi)$ as:
\begin{equation}
    \label{beta_approx_varphi}
    \beta(\varphi ) \sim - \frac{\mathrm{d} \ln V(\varphi)}{\mathrm{d} \varphi} =  - 2 \frac{\tilde{f}_{,\varphi}(\varphi)}{\tilde{f}(\varphi) \left[1 + \xi \tilde{f}(\varphi) \right]} ,
\end{equation}  
where $\tilde{f}(\varphi) \equiv f(\phi(\varphi))$. Characterizing the system in terms of $\varphi$ helps to have a deeper comprehension of this model and leads to an interpretation of the attractor at strong coupling. In the rest of this section we discuss the limits of large and small $\xi$ and we present an explicit example to understand the interpolation between these two regimes.

\subsection{Strong and weak coupling limits.}
\label{sec:strong}
In the strong coupling limit we have $1 \ll \xi $ and thus the lowest order approximation for \eqref{beta_approx_varphi} simply reads:
\begin{equation}
  \label{beta_strong}
    \beta(\varphi ) \simeq  - \frac{2}{\xi} \frac{\tilde{f}_{,\varphi} (\varphi )}{ \tilde{f}^2 (\varphi)}  .  
\end{equation} 
Using eq.\eqref{generalF} we can get the lowest order expression for $F(\phi)$ i.e.
\begin{equation}
  \label{F_strong}
  F(\phi) \simeq \frac{3}{2}\left( \frac{f_{,\phi} (\phi)}{ f(\phi)} \right)^2. 
\end{equation}
We can substitute eq.\eqref{F_strong} into eq.\eqref{def_varphi} and integrate to get:
\begin{equation}
\label{def_varphi_strong}
f(\phi(\varphi)) = \tilde{f}(\varphi) = f_{\mathrm{f}}\exp \left[  \sqrt{\frac{2}{3}} (\varphi - \varphi_{\mathrm{f}} ) \right],
\end{equation}
where we defined $f_{\mathrm{f}} \equiv \tilde{f}(\varphi_{\mathrm{f}})$. It is crucial to notice that in the limit $1 \ll \xi$ the expression for $f(\varphi)$ does not depend on the explicit choice for $f(\phi)$! As shown in eq.\eqref{beta_strong}, the expression of $\beta(\varphi)$ in the limit of a strong coupling is only depending on $\tilde{f}(\varphi)$. We can thus substitute eq.\eqref{def_varphi_strong} into eq.\eqref{beta_strong} to get:
\begin{equation}
\label{explicit_beta_strong}
\beta(\varphi) = -\sqrt{\frac{8}{3}} \frac{1}{\xi f_{\mathrm{f}}} \exp \left[ - \sqrt{\frac{2}{3}} (\varphi - \varphi_{\mathrm{f}} ) \right].
\end{equation}
It is important to stress that at the end of inflation $1 + p/\rho $ is close to one and thus eq.\eqref{eq_of_state} implies that $\left| \beta(\varphi_{\mathrm{f}}) \right| \sim 1 $. Eq.\eqref{explicit_beta_strong} then leads to $f_{\mathrm{f}} \sim \sqrt{8/3} / \xi$ that can be substituted into eq.\eqref{explicit_beta_strong} to conclude that: 
\begin{equation}
\label{beta_strong_ff}
\beta(\varphi) = -\exp \left[ - \sqrt{\frac{2}{3}} (\varphi - \varphi_{\mathrm{f}} ) \right].
\end{equation}
It is then clear that the expression for $\beta(\varphi)$, in the limit of big $\xi$, is independent on the explicit choice for $f(\phi)$. As the dynamics of the system during the inflationary phase is completely specified by $\beta(\varphi)$, this directly leads to the universality. In particular, we notice that $\beta(\varphi)$ approaches the exponential class of \cite{Binetruy:2014zya}. This universality class is entirely determined by a single critical exponent, denoted with $\gamma$, that in this case is equal to the $\sqrt{2 /3 }$ factor in the exponential of \eqref{beta_strong_ff}. As discussed in \cite{Binetruy:2014zya}, the scalar spectral index and the tensor to scalar ratio are given by: 
        \begin{eqnarray} 
          \label{ns_strong}
            n_{s} - 1 &\simeq& - \frac{2}{N}, \\
            \label{r_strong}
            r  &\simeq& \frac{8}{\gamma^2 N^2} = \frac{12}{N^2} .
    \end{eqnarray}
As expected, these results are in perfect agreement with the ones discussed in \cite{Kallosh:2013tua}. In this framework, the independence of $\beta(\varphi)$ on $f(\phi)$ directly leads to the universality for the values of $n_s$ and $r$. In particular, in terms of the $\beta$-function formalism, the appearence of a universal attractor at strong coupling corresponds to the flow of the system into a particular universality class. \\

For completeness we can also express $N(\varphi)$ as:
\begin{equation}
\label{efold_ff}
N(\varphi) =  - \int_{\varphi_f}^{\varphi} \frac{1}{\beta(\hat{\varphi})} d\hat{\varphi} =  \sqrt{ \frac{3}{2}}  \left\{ \exp \left[  \sqrt{\frac{2}{3}} \left( \varphi - \varphi_{\mathrm{f}} \right)     \right]  - 1 \right\} .
\end{equation}
Substituting eq. \eqref{efold_ff} into eq. \eqref{beta_strong_ff} we find that choosing values for $N$ in the range $ [50,60]$ we get $\beta \in [-0.024,-0.02]$. Notice that in the limit of strong coupling, the dynamics in terms of $\varphi$ does not depend on $\xi$. It is also interesting to point out that in this case the asymptotic fixed point is reached for $N \rightarrow \infty$ that corresponds to $\varphi \rightarrow \infty$. \\

It is interesting to point out that $\xi = 0$ corresponds to a minimal coupling between the inflaton and gravity. In this case we can again use the equations derived in Sec.\ref{sec:beta_function}, but the expression for $\tilde{f}(\varphi)$ given by eq.\eqref{def_varphi_strong} does not hold. As a consequence we are not expecting to obtain a model independent expression for $\beta(\varphi)$ and thus results will be model dependent. In particular, by choosing particular parametrizations for $\beta(\varphi)$, we can reproduce the universality classes introduced in \cite{Binetruy:2014zya}. As in the limit of a weak coupling we are just introducing a small variation with respect to the case of $\xi = 0$, we will only obtain a little departure from the standard results. In particular it is possible to prove that in the limit of a weak coupling the lowest order expressions for $n_s$ and $r$ correspond to the ones presented in \cite{Kallosh:2013tua}.

\subsection{An explicit example.}
\label{sec:example}
\begin{figure}[htb!]
\centering

{\includegraphics[width=1 \columnwidth]{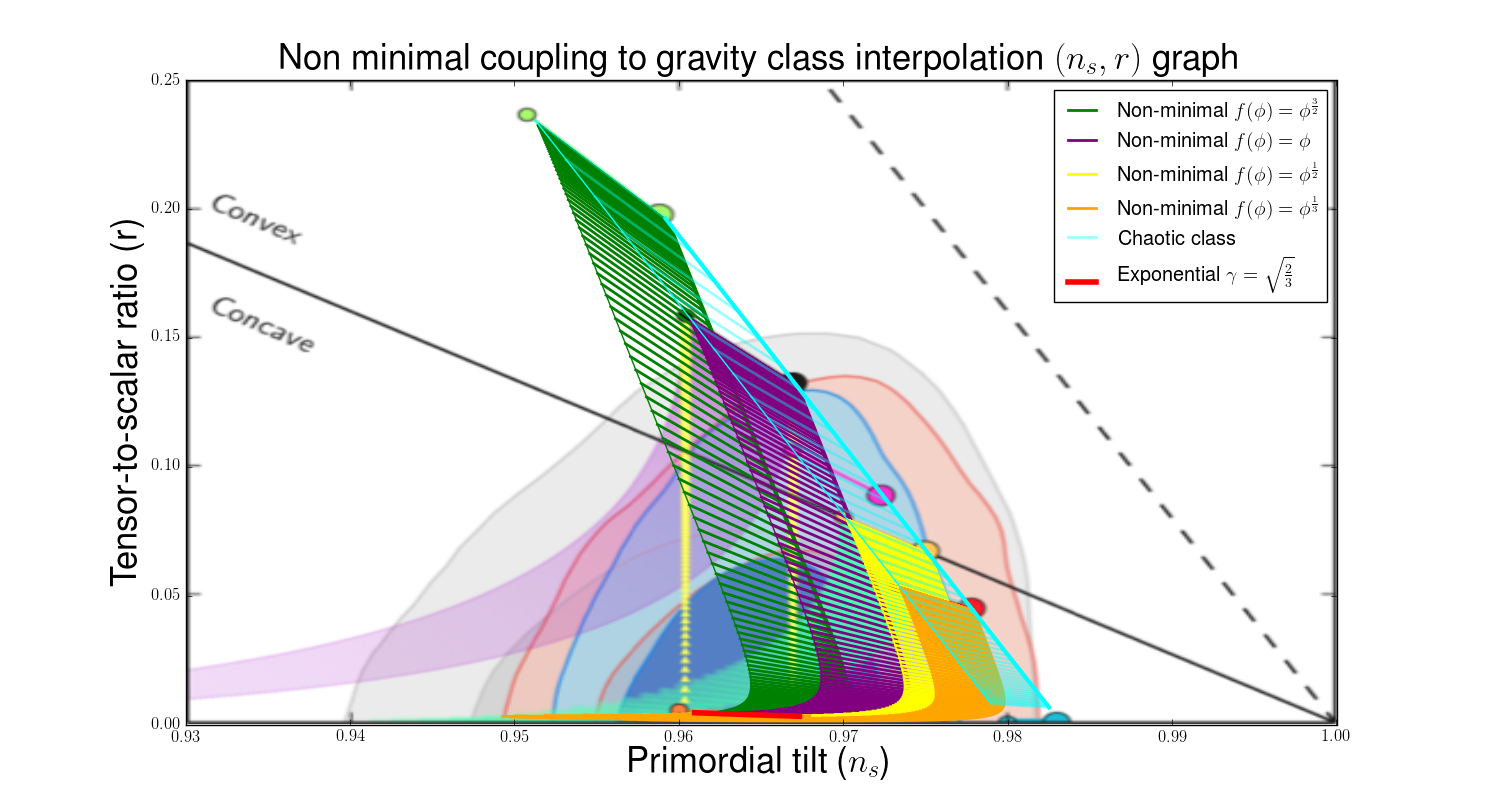}}\\
\caption{   \label{figure1} \footnotesize Numerical predictions for $n_s$ and $r$ for the non-minimally coupled models are compared with Chaotic class with some values for $\alpha$ in the range $[0.1,3]$ and exponential class with $\gamma = \sqrt{2/3}$. The results are presented with the famous Planck $(n_s,r)$ graph as a background \cite{Ade:2015lrj}. In particula we have Planck 2013 (grey contours), Planck TT+lowP (red contours), and Planck TT,TE,EE+lowP (blue contours). 
}
\end{figure}

\begin{figure}[htb!]
\centering

{\includegraphics[width=0.95 \columnwidth]{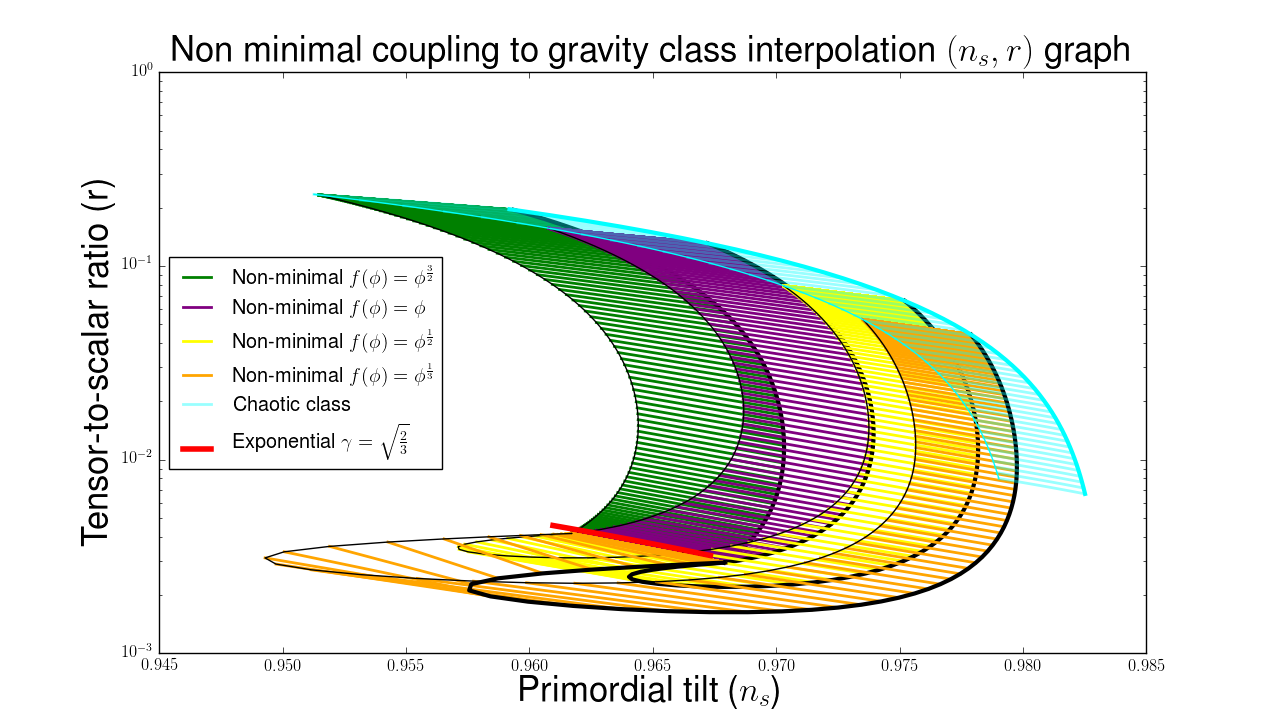}}\\
\caption{   \label{figure2}\footnotesize  Numerical predictions of Fig.\ref{figure1} are presented in a semilogarithmic plot.}
\end{figure}

In this section we present an example to be have a better understanding of the transition from the weak to the strong coupling limits. In particular, we consider some particular models by specifying an explicit expression for $f(\phi)$. From eq.\eqref{beta_varphi} and eq.\eqref{potential_1}, it should be clear that $\beta(\varphi) \sim - 2\alpha/\varphi$ simply gives:
\begin{equation}
W(\varphi) = W_\textit{f} \left (\frac{\varphi}{\varphi_\textit{f}} \right)^{\alpha}, \qquad \qquad V(\varphi)  = \frac{3 W_\textit{f}^2}{4} \left (\frac{\varphi}{\varphi_\textit{f}} \right)^{2\alpha} = V_\textit{f} \left (\frac{\varphi}{\varphi_\textit{f}} \right)^{2\alpha}.
\end{equation}
This clearly corresponds to the well known case of chaotic inflation \cite{Linde:1983gd}. As in the limit of small $\xi$ we have $\varphi \sim \phi$ and $\beta(\varphi) \sim - 2 \tilde{f}_{,\varphi}(\varphi)/\tilde{f}(\varphi)$, to obtain this expression for $\beta(\varphi)$ we simply choose $f(\phi) = \phi^\alpha$. It is well known that in this case the lowest order predictions for the $n_s$, scalar spectral index, and $r$, tensor to scalar ratio, are given by:
  \begin{equation} 
  \label{chaotic_predictions}
    n_{s} \simeq 1 - \frac{1+\alpha}{N}, \qquad \qquad              r  \simeq \frac{8\alpha}{N}.
\end{equation}
On the contrary, the strong limit predictions have been discussed in Sec \ref{sec:strong}, and these are given by eq.\eqref{ns_strong} and eq.\eqref{r_strong}. Variating the value of $\xi$, we expect to shift from the model dependent regime to the universal attractor at strong coupling. Numerical results for our choice for $f(\phi)$ are shown in Fig. \ref{figure1} and Fig. \ref{figure2}. In this particular case the parametrization of $\beta(\varphi)$ is completely specified by the value of the critical exponent $\alpha$. Once this constant is fixed, we can compute numerical predictions as a function of $N$, number of e-foldings. in Fig. \ref{figure1} and Fig. \ref{figure2} we use different colors to plot models associated with a different values for $\alpha$. The solid black lines in the plot of Fig. \ref{figure2} are used to follow the variation of $\xi$ while the values of $\alpha$ and $N$ are fixed. The thick line corresponds to $ N = 60 $ and the thin one corresponds to $N = 50$. Numerical results are compared with the ones obtained for the chaotic class i.e. $\beta(\varphi) = - \alpha / \varphi$ with some values for $\alpha$ in the range $[0.1,3]$ and for the exponential class i.e. $\beta(\varphi) = - \exp \left[ - \gamma \varphi \right]$ with $\gamma = \sqrt{2/3}$. In limit of a weak coupling numerical predictions match with the chaotic class while in the strong coupling limit we approach the predicted exponential class attractor. In the intermediate region we have a whole set of valid inflationary models that actually interpolate between the two fundamental classes.\\

This behavior is quite similar to the mechanism of interpolation discussed in \cite{Binetruy:2014zya}. In these paper we have shown that, by introducing a new scale $f$, it is possible to construct a $\beta$-function that approaches different universality classes as we consider different values for $f$. In particular we have shown that a small and a large field regime can be reached. For example let us consider a model with a scalar field $\chi$ and let us assume that $\beta(\chi) = g(\chi)$ is the $\beta$-function for this model. As we are interested in studying an inflationary stage, the system is close to zero of the $\beta$-function. Without loss of generality we assume that $\beta(\chi = 0) = 0$. Finally we consider a model with $\beta$-function defined by $\beta(\chi) = \epsilon g(\chi)$ where $\epsilon \ll 1$ is a constant. It should be clear that this system inflates for all the values for $\chi$ such that $\beta(\chi) = \epsilon g(\chi) \ll 1$ and thus we can have inflation even for $g(\chi) \sim 1$. As a matter of fact the small field regime is stretched and it can be reached even for bigger values for $\chi$. In the case of a scalar field with a non-minimally coupling with gravity the role of the scale $f$ appears to be played by the coupling $\xi$. In particular this can be shown expressing the $\beta$-function in terms of $\phi$:
\begin{equation}
\label{beta_phi}
 \bar{\beta}(\phi) = \beta(\varphi(\phi)) = - \left( \frac{\mathrm{d} \phi}{\mathrm{d} \varphi} \right) \frac{\mathrm{d} \ln \bar{V}(\phi)}{\mathrm{d} \phi}.
\end{equation} 
Usign eq.\eqref{generalF} and eq.\eqref{F_strong} we express the $\beta$-function in the limit of strong coupling as:
\begin{equation}
\label{final_beta_phi}
 \bar{\beta}(\phi) = - \sqrt{\frac{8}{3}} \frac{\phi^{-\alpha}}{\xi}.
\end{equation} 
It should be clear that a zero of this function is reached for $1 \ll \phi^\alpha$. However, by choosing a large value for $\xi$, it is still possible to have inflation in the limit of $\phi^\alpha \ll 1$.
In particular, consistently with \cite{Kallosh:2013tua}, this mechanism allows the production of cosmological perturbations in the regime $ \phi^\alpha \ll 1$. By choosing a large value of $\xi$ we have thus stretched the asymptotic large field regime so that it can be obtained even for small values of $\phi$.

\section{A more general discussion on non-minimal coupling}
\label{sec:general_case}
In this section we are interested in discussing a more general parametrization for the model for a scalar field with a non minimal coupling with gravity. In particular we follow the proposal of \cite{Kallosh:2013tua} and we start by considering the same lagrangian of eq.\eqref{minimal-action} i.e.
 \begin{equation}
  \label{non_minimal_action_2}
    S =\int\mathrm{d}^4x\sqrt{-g}\left( -  \frac{\Omega(\phi)}{2\kappa^2}R +  X - V_J(\phi) \right),
  \end{equation}
but we introduce an additional functional freedom in the model i.e. 
\begin{equation}
\Omega(\phi) = 1 + \xi f(\phi) ,\qquad \qquad \qquad V_J(\phi) = \lambda^2 g^2(\phi) ,
\end{equation}
where both $f(\phi)$ and $g(\phi)$ are generic functions of $\phi$. As argued in \cite{Kallosh:2013tua}, by studying the system in terms of $\phi$, it seems reasonable to assume that small variations of the potential should not affect the occurrence of the attractor. In the following we will prove that in general this conclusion does not appear to hold.\\

By means of a conformal transformation we recover the Einstein frame action of eq.\eqref{non-minimal-action}. In this case the expressions for $\Omega(\phi)$ and $V(\phi)$ are given by:
\begin{equation}
\label{omega_V}
\Omega(\phi) = 1 + \xi f(\phi) , \qquad \qquad \bar{V}(\phi) = \frac{\lambda^2 g^2(\phi) }{\Omega^2}.
\end{equation}
Following the procedure defined in Sec.\ref{sec:beta_function}, we describe the system in terms of a new field $\varphi$ with a canonically normalized kinetic term. Substituting eq.\eqref{generalF} into eq.\eqref{def_varphi} we get: 
\begin{equation}
\label{general_vaphi}
\left( \frac{\mathrm{d} \varphi}{\mathrm{d} \phi} \right)^2  = F(\phi) = \frac{1 + \xi f + \frac{3}{2} \xi^2 f^2_{,\phi} }{(1 + \xi f(\phi))^2}.
\end{equation}
The expression for the $\beta$-function associated with the system reads: 
\begin{equation}
\label{beta_general}
\beta(\varphi) \sim - \frac{\mathrm{d} \ln V(\varphi)}{\mathrm{d} \varphi} = - 2 \left( \frac{\tilde{g}_{,\varphi}(\varphi)}{\tilde{g}(\varphi)} - \frac{\xi \tilde{f}_{,\varphi}(\varphi)}{1 + \xi \tilde{f}(\varphi)} \right), 
\end{equation}
where in analogy with $\tilde{f}(\varphi)$, we defined $\tilde{g}(\varphi) = g(\phi(\varphi))$. Without loss of generality we can parametrizations $\tilde{g}(\phi)$ as: 
\begin{equation}
\label{eq_parametization}
\tilde{g}(\varphi) = \tilde{f}(\varphi) \tilde{h}(\varphi),
\end{equation}
where $\tilde{h}(\varphi)$ is a generic function of $\varphi$. It is important to stress that we are not specifying an explicit expression for $\tilde{h}(\varphi)$ and thus we can produce a quite general description of the problem. We can substitute eq.\eqref{eq_parametization} into eq.\eqref{beta_general} to get:
\begin{equation}
\label{beta_general_2}
\beta(\varphi) = - 2 \left\{ \frac{\tilde{h}_{,\varphi}(\varphi)}{\tilde{h}(\varphi)} +  \frac{ \tilde{f}_{,\varphi}(\varphi)}{\tilde{f}(\varphi) \left[ 1 + \xi \tilde{f}(\varphi) \right]} \right\}.
\end{equation}
It is easy to notice that in the case of $\tilde{h}_{,\varphi}(\varphi)/\tilde{h}(\varphi) = 0$, this equation is exactly equal to eq.\eqref{beta_approx_varphi}. In particular, in the limit of strong coupling $\xi$ eq.\eqref{beta_general_2} simply reads:
\begin{equation}
\label{beta_general_3}
\beta(\varphi) = - 2 \left[ \frac{\tilde{h}_{,\varphi}(\varphi)}{\tilde{h}(\varphi)} +  \frac{ \tilde{f}_{,\varphi}(\varphi)}{ \xi \tilde{f}^2(\varphi) } \right].
\end{equation}
It is interesting to point out that choosing $\tilde{f}(\varphi) = \tilde{g}(\varphi)$ or equivalently $\tilde{h}(\varphi) = 1$, the zero order term in $1/\xi$ is set equal to zero. Under this assumption the expression for $\beta(\varphi)$ is thus dominated by the first order term $1/\xi$. In particular the $\beta$-function is simply given by eq.\eqref{beta_strong} and thus we reduce to the case discussed in Sec. \ref{sec:strong}. Relaxing the assumption of $\tilde{h}(\varphi) = 1$, we can consider the case of a zero order term different from zero. As an inflationary stage corresponds to $\beta(\varphi) \rightarrow 0$ any choice of $\tilde{h}(\varphi)$ that satisfies: 
\begin{equation}
\label{eq_evasion}
 \frac{ \tilde{f}_{,\varphi}(\varphi)}{ \xi \tilde{f}^2(\varphi) } \ll \frac{\tilde{h}_{,\varphi}(\varphi)}{\tilde{h}(\varphi)} \rightarrow 0,
\end{equation}
describing a viable inflationary model. As no other restriction has been imposed on the choice for $\tilde{h}(\varphi)$, we can immediately conclude that in general the attractor at strong coupling can be evaded. In the Sec. \ref{sec:General_example} we present an explicit example to discuss the conditions to preserve the attractor at strong coupling. In Sec. \ref{sec:Further_generalizations} we show that models defined via further generalizations of the action eq.\eqref{non_minimal_action_2} are still included in this class and we investigate the characterization of the $\alpha$-attractors of Kallosh and Linde \cite{Kallosh:2013hoa,Kallosh:2013daa,Kallosh:2013yoa,Kallosh:2015lwa} in terms of our formalism. Some other examples of the parametrization for $\tilde{h}(\varphi)$ are discussed in Appendix \ref{Appendix}.

\subsection{Polynomial expansion.}
\label{sec:General_example}
Let us assume that $f(\phi)$ and $g(\phi)$ admit a Taylor expansion in terms of $\phi$:
\begin{equation}
\label{Taylor}
f(\phi) = \sum_{i = 0}^\infty f_i \phi^i, \qquad \qquad g(\phi) = \sum_{i = 0}^\infty g_i \phi^i .
\end{equation}
Let us restrict to the case of both $f(\phi)$ and $g(\phi)$ vanishing for a certain value of $\phi$. By means of a field redefinition we can fix $f_0 = g_0 = 0$. Without loss of generality we can also rescale $\lambda $ and $\xi$ to impose $f_1 = g_1 = 1$. The case $f(\phi) = g(\phi)$ has been discussed in Sec. \ref{sec:strong}, and in particular we have shown that under this assumption it is possible to choose $\xi$ such that $\phi \ll 1$. As the first order terms of eq.\eqref{Taylor} are imposed to be equal, and high order orders in terms of $\phi$ are expected to be negligible,  it would be reasonable to conclude that the attractor at strong coupling should be preserved. Surprisingly, expressing the dynamics in terms of $\varphi$, it is possible to show that the attractor at strong coupling may be evaded! Let us fix a particular expression for $f(\phi)$ and $g(\phi)$, in particular we choose:
 \begin{equation}
\label{explicit_expansion}
f(\phi) = \phi, \qquad \qquad g(\phi) =\phi + g_{n+1} \phi^{n+1} = \phi (1 +g_{n+1} \phi^{n} ) .
\end{equation}
In the strong coupling limit, the lowest order approximation for eq.\eqref{general_vaphi} simply reads:
\begin{equation}
  \label{example_beta_strong}
    \left( \frac{ \textit{d} \varphi}{ \textit{d} \phi} \right)^2 =  F(\phi) \simeq \frac{3}{2}\left( \frac{f_{,\phi} (\phi)}{ f(\phi)} \right)^2.  
\end{equation} 
We can integrate eq.\eqref{example_beta_strong} to get an explicit expression for $\tilde{f}(\varphi)$:
\begin{equation}
\label{example_varphi}
\tilde{f}(\varphi) = \tilde{f}_{\mathrm{f}}\exp \left[ \sqrt{\frac{2}{3}} (\varphi - \varphi_{\mathrm{f}} ) \right].
\end{equation} 
Finally we substitute into eq.\eqref{explicit_expansion} to get:
\begin{eqnarray}
\label{example_phi_varphi}
\phi &=& f(\phi) = \tilde{f}(\varphi) = \tilde{f}_{\mathrm{f}}\exp \left[ \sqrt{\frac{2}{3}} (\varphi - \varphi_{\mathrm{f}} ) \right], \\
\label{explicit_g}
\tilde{g} (\varphi) &=& \tilde{f}_{\mathrm{f}}\exp \left[  \sqrt{\frac{2}{3}} (\varphi - \varphi_{\mathrm{f}} ) \right] \left\{ 1 + g_{n+1} \tilde{f}_{\mathrm{f}}^{\ n} \exp \left[  \sqrt{\frac{2}{3}} n (\varphi - \varphi_{\mathrm{f}} ) \right] \right\},
\end{eqnarray}
where $\tilde{g} (\varphi)  = g(\phi(\varphi))$. It should be clear that this corresponds to: 
\begin{equation}
\label{example_h}
 \tilde{h}(\varphi) =  1 + g_{n+1} \tilde{f}_{\mathrm{f}}^{\ n} \exp \left[  \sqrt{\frac{2}{3}} n (\varphi - \varphi_{\mathrm{f}} ) \right].
\end{equation}
Using eq.\eqref{beta_general_2} we can then compute the explicit expression for $\beta(\varphi)$: 
\begin{equation}
\label{explicit_beta}
 \beta(\varphi) = -\sqrt{\frac{8}{3}}  \left\{ \frac{n g_{n+1} \tilde{f}^{\ n}_{\mathrm{f}} \exp \left[  \sqrt{\frac{2}{3}} n (\varphi - \varphi_{\mathrm{f}} ) \right]  }  {1 + g_{n+1} \tilde{f}^{\ n}_{\mathrm{f}}\exp \left[  \sqrt{\frac{2}{3}} n (\varphi - \varphi_{\mathrm{f}} ) \right] }  +  \frac{1  }{1 + \xi  \tilde{f}_{\mathrm{f}}\exp \left[  \sqrt{\frac{2}{3}} (\varphi - \varphi_{\mathrm{f}} ) \right] } \right\}.
\end{equation}
Notice that the first term on the right hand side of eq.\eqref{explicit_beta} gives the zero order contribution in $1/\xi$ while the second term on the right hand side of eq.\eqref{explicit_beta} is a first order term in $1/\xi$. It is important to stress that imposing $ g_{n+1} = 0$, is equivalent to fix $f(\phi) = g(\phi)$. As discussed in Sec.\ref{sec:strong}, in this case the inflationary regime is reached for large positive values for $\varphi$ and $\beta(\varphi)$ is approximated by eq.\eqref{beta_strong_ff}. On the contrary when $ g_{n+1} \neq 0$, the second term on the right hand side of eq.\eqref{explicit_beta} is negligible with respect to the first one\footnote{The consistency of this assumption is discussed in the following.}. Under this assumption $\beta(\varphi)$ can be approximated as:
\begin{equation}
\label{explicit_beta_2}
 \beta(\varphi) \sim -\sqrt{\frac{8}{3}}  \left\{ \frac{ n g_{n+1} \tilde{f}^{\ n}_{\mathrm{f}} \exp \left[  \sqrt{\frac{2}{3}} n (\varphi - \varphi_{\mathrm{f}} ) \right]  }  {1 + g_{n+1} \tilde{f}^{\ n}_{\mathrm{f}}\exp \left[  \sqrt{\frac{2}{3}} n (\varphi - \varphi_{\mathrm{f}} ) \right] } \right\}.
\end{equation}
In this case the zero of $\beta(\varphi)$ that corresponds to the inflationary phase is thus reached for large negative values for $\varphi$. In this regime the expressions for $\beta(\varphi)$ and $N(\varphi)$ are:
\begin{eqnarray}
\label{explicit_beta_3}
 \beta(\varphi) &\sim& -\sqrt{\frac{8}{3}} n g_{n+1} \tilde{f}^{\ n}_{\mathrm{f}} \exp \left[  \sqrt{\frac{2}{3}} n (\varphi - \varphi_{\mathrm{f}} ) \right] ,  \\
\label{efold_explicit}
N(\varphi) &=& - \frac{3}{4} \frac{1}{n^2 g_{n+1} \tilde{f}^{\ n}_{\mathrm{f}}}  \left\{ \exp \left[ - \sqrt{\frac{2}{3}} n \left( \varphi - \varphi_{\mathrm{f}} \right)     \right]  - 1 \right\}.
\end{eqnarray}
To ensure $0 \leq N(\varphi)$, we impose $0 < - f^n_{\mathrm{f}} g_{n+1}$. Following the same procedure of Sec.\ref{sec:strong}, we also impose the condition $\left| \beta(\varphi_{\mathrm{f}}) \right| \sim 1 $ to fix the value of $\beta(\varphi)$ at the end of inflation: 
\begin{equation}
\label{condition}
 \left| \beta(\varphi_{\mathrm{f}}) \right| = \left| \sqrt{\frac{8}{3}} n g_{n+1} \tilde{f}^{\ n}_{\mathrm{f}}  \right| \sim 1.
\end{equation}
As $n$ and $g_{n+1}$ are expected to be of order one, we can conclude that $ f_{\mathrm{f}}$ is expected to be of order one too. Finally, using eq.\eqref{condition} we express $\beta(\varphi)$ and $N(\varphi)$ as:
\begin{eqnarray}
\label{explicit_beta_final}
 \beta(\varphi) &\sim& \exp \left[  \sqrt{\frac{2}{3}} n (\varphi - \varphi_{\mathrm{f}} ) \right], \\
 \label{explicit_efold_final}
 N(\varphi) &=& \sqrt{\frac{3}{2 n^4}} \left\{ \exp \left[ - \sqrt{\frac{2}{3}} n \left( \varphi - \varphi_{\mathrm{f}} \right)     \right]  - 1 \right\}.
\end{eqnarray}
The expression for $\beta(\varphi)$, in the limit of big $\xi$, thus depends on $n$ and this leads to the evasion from the univerality. In particular, $\beta(\varphi)$ approaches the exponential class of \cite{Binetruy:2014zya} with $\gamma = n \sqrt{2 /3 }$. The corrisponding expression for the scalar spectral index and for the tensor to scalar ratio are given by: 
\begin{equation}
\label{ns_r_evaded_attractor}
n_s = 1 - \frac{2}{N}, \qquad \qquad \qquad r = \frac{12}{n^2 N^2}.
\end{equation}
It is interesting to notice that the attractor at strong coupling of \cite{Kallosh:2013tua} can be reproduced by imposing $n = 1$. Actually it is possible to go further and prove that the attractor can be recovered under some more general condition. As argued during this section, the inflationary phase is reached for $\varphi \ll -1$ and this corresponds to $\phi \ll 1$. In this regime higher order corrections to the expression of $g(\phi)$:
 \begin{equation}
\label{generalized_expansion}
f(\phi) = \phi, \qquad \qquad g(\phi) =\phi + g_{n+1} \phi^{n+1} + \sum_{i = n+2 }^\infty g_i \phi^i  ,
\end{equation}
are not producing significat changes in the lowest order expressions for $\beta(\varphi)$ and $N(\varphi)$ given in eqs.\eqref{explicit_beta_final},\eqref{explicit_efold_final}. In particular this implies that assuming $g_2 \neq 0$, the dominant contribution to the expression of $\beta(\varphi)$ is fixed by the term with $n=1$. This condition is thus sufficient to preserve the attractor\footnote{Notice that this is a specific feature of the parametrization of eq. \eqref{generalized_expansion}. As discussed in Sec. \ref{sec:general_case}, the attractor can be evaded under the quite general condition of eq.\eqref{eq_evasion}. Some explicit examples of the evasion are presented in the appendix \ref{Appendix}.}. Conversely, other attractors are find for different values of $1<n$. \\

To conclude this section we discuss the consistency of the assumption that the second term on the right hand side of eq.\eqref{explicit_beta} is negligible with respect to the first one. To be sure that this term is subdominant from the end of inflation up to the production of cosmological perturbations we need: 
\begin{equation}
\label{attractor_condition}
  \frac{1  }{1 + \xi  \tilde{f}_{\mathrm{f}}\exp \left[  \sqrt{\frac{2}{3}} (\varphi_{\mathrm{H}} - \varphi_{\mathrm{f}} ) \right] }  \ll  \exp \left[  \sqrt{\frac{2}{3}} n (\varphi_{\mathrm{H}} - \varphi_{\mathrm{f}} ) \right]   \ll 1,
\end{equation} 
where $\varphi_{\mathrm{H}}$ is the value of $\varphi$ at the production of cosmological perturbation. Using the expression for $N(\varphi)$ given by eq.\eqref{explicit_efold_final} it is clear that eq.\eqref{attractor_condition} satisied if $N_{\mathrm{H}}^2/\xi \ll 1$.\\

\subsection{$\alpha$-attractors.}
\label{sec:Further_generalizations}
It is interesting to notice that the class of models described in this section also includes further generalizations of the lagrangian of eq.\eqref{non_minimal_action_2}. In particular some of these generalizations have been presented in \cite{Galante:2014ifa} and \cite{Kallosh:2014laa}. Following the proposal of \cite{Galante:2014ifa} we consider the general Jordan frame action\footnote{$\kappa^2$ is set equal to 1.} to describe a homogeneous scalar field with a non-minimal coupling with gravity non-minimally:
  \begin{equation}
  \label{general_action_jordan}
    S=\int\mathrm{d}^4x\sqrt{-g}\left( -  \Omega(\phi)\frac{R}{2} + K_J(\phi) X - V_J(\phi) \right).
  \end{equation}
As usual we perform a conformal transformation:
  \begin{equation}
\label{conformal}
  g_{\mu\nu} \rightarrow \Omega(\phi)^{ -1} g_{\mu\nu},
  \end{equation}
to get the Einstein frame formulation of the theory: 
    \begin{equation}
  \label{general_action_einstein}
    \mathcal{L}_E = - \frac{R}{2} +  F(\phi)  X - V(\phi) ,
  \end{equation}
  where we defined $F(\phi)$ and $V(\phi)$ as:
  \begin{equation}
  F(\phi) \equiv \left[ \frac{K_J (\phi)}{\Omega(\phi)} + \frac{3}{2} \left(  \frac{\textrm{d} \ln \Omega(\phi)}{\textrm{d}\phi}\right)^2 \right] \qquad \qquad V(\phi) = \frac{V_J(\phi)}{\Omega^2(\phi)}.
  \end{equation}
It is clear that the cases discussed in the previous sections can be recovered simply imposing $K_J (\phi) = 1$. Again we can define a new field $\varphi$:
    \begin{equation}
  \label{general_varphi_def}
    \left(  \frac{\textrm{d} \varphi}{\textrm{d}\phi}\right)^2  \equiv F(\phi) = \left[ \frac{K_J (\phi)}{\Omega(\phi)} + \frac{3}{2} \left(  \frac{\textrm{d} \ln \Omega(\phi)}{\textrm{d}\phi}\right)^2 \right] ,
  \end{equation}
that has a canonically normalized standard kinetic term. In particular the lagrangian for this field simply reads:
    \begin{equation}
  \label{general_varphi_einstein}
    \mathcal{L}_E = - \frac{R}{2} + \frac{(\partial \varphi)^2}{2} - \tilde{V}(\varphi) ,
  \end{equation}
  where $\tilde{V}(\varphi) $ is defined as $\tilde{V}(\varphi) = V(\phi(\varphi))$. As in terms of the canonically normalized field $\varphi$ the three functional dependece are merged into $\tilde{V}(\varphi)$, the model construction reduces to fixing a particular parametrization for this function. Usign eq.\eqref{beta_varphi}, and the lowest order approximation $V(\varphi) \sim \frac{3}{4} W^2(\varphi)$ we can finally express the $\beta$-function as:
    \begin{equation}
  \label{beta_varphi_def}
   \beta(\varphi) \sim -  \frac{\textrm{d} \ln \tilde{V}(\varphi)}{\textrm{d}\varphi} .
  \end{equation}
Again the whole dynamics of the model is thus fixed by the parametrization of the $\beta$-function. As different choices for $\Omega(\phi), K_J(\phi)$ and $V_J(\phi)$ lead to the same expression for $\beta$, this explains the possibility for degeneracies to arise.\\

Several models described by the action of eq.\eqref{general_action_jordan} has been presented in \cite{Galante:2014ifa} and \cite{Kallosh:2014laa}. In this paper we consider the $\alpha$-attractors of \cite{Kallosh:2015lwa} as an interesting example for this class of models. In particular let us consider the case of \emph{T-models} \cite{Kallosh:2013hoa}. \emph{T-models} can be described in terms of the action of eq.\eqref{general_action_jordan} by fixing:  
\begin{equation}
\label{alpha_attrators}
  \left(  \frac{\textrm{d} \varphi}{\textrm{d}\phi}\right)^2 = F(\phi) = \left( 1 - \frac{\phi^2}{6 \alpha}\right)^{-2} \qquad \qquad \qquad V(\phi) = \frac{m^2}{2} \phi^2.
  \end{equation}
Using eq.\eqref{alpha_attrators} we define the canonically normalized field and using eq.\eqref{omega_V} and eq.\eqref{eq_parametization} we can compute the explicit expression for $h(\phi)$. In particular we get:
 \begin{equation}
\phi = \sqrt{6\alpha} \tanh\left(\frac{\varphi}{\sqrt{6\alpha}} \right), \qquad \qquad \qquad h(\phi) \sim \phi.
  \end{equation}
Finally we can use eq.\eqref{beta_general_3} to compute the expplicit expression for the $\beta$-function:
 \begin{equation}
 \label{beta_alpha_attractors}
 \beta(\varphi) = - \sqrt{\frac{2}{3 \alpha}} \left[ \frac{1 - \tanh^2 \left( \frac{\varphi}{\sqrt{6 \alpha}} \right)}{\tanh \left( \frac{\varphi}{\sqrt{6 \alpha}} \right)} \right] \sim -  \exp \left[ - \sqrt{\frac{2}{3 \alpha}} ( \varphi-\varphi_f) \right].
  \end{equation}
Eq.\eqref{beta_alpha_attractors} imples that the $\beta$ function for \emph{T-models} falls in the exponential class of \cite{Binetruy:2014zya}. As already discussed in this paper, the predictions for $n_s$ and $r$ are thus given by:
\begin{equation}
\label{ns_r_alpha_attractor}
n_s = 1 - \frac{2}{N}, \qquad \qquad \qquad r = \frac{8}{\gamma^2 N^2} = \frac{12 \alpha}{N^2}.
\end{equation}

Similar conclusions can be draw for the other models for $\alpha$-attractors presented in \cite{Kallosh:2013hoa}.

\section{Conclusions}
\label{sec:conclusions}

In the analysis of this paper, by means of a conformal transformation and of a field redefinition, we have discussed the problem of inflationary models with a non-minimal coupling with gravity in terms of a single field with a canonically normalized kinetic term. In particular we have shown, the application of the $\beta-$function formalism, helps to understand the asymptotic behavior of the system during the inflationary phase. In this framework, the fall of the system into the attractor is interpreted as the approach of a universality class. In this sense, the formulation of the problem in this framework, should not be seen as a simple rewriting of the results obtained with standard methods, but on the contrary it should be considered as a further generalization.\\

The $\beta-$function formalism appears to be extremely useful when we investigate the stability of the attractor at strong coupling under generalizations of the theory. In particular, once we have defined the $\beta-$function associated with our system, it has been easy to identify the dominant contribution to characterize inflation.
Specifically, in Sec.\ref{sec:general_case}, we have discussed the possibility of introducing an additional functional freedom in the model. In this case the behavior of the system is dominated by the zeroth order term that conversely was set equal to zero in the treatment followed in Sec.\ref{sec:beta_function}. As in general this term can be chosen arbitrarily, it leads to the possibility of evading the attractor at strong coupling. A critical review of the conditions required to preserve the attractor at strong coupling has been presented and the existence of different attractors has been shown. \\

The further generalization discussed in \cite{Galante:2014ifa} and \cite{Kallosh:2014laa} have been presented. In these works it was shown that a slight modification of the theory may lead to the existence of other attractors. Indeed for these models an analogous of the treatment presented in this paper can be carried out and it leads to similar conclusions. In particular we have presented the application of our formalism to the case of the $\alpha$-attractors of \cite{Kallosh:2015lwa}. A further generalization of the formalism proposed in \cite{Binetruy:2014zya} can be also useful to have a deeper understanding of more general models with a non-standard kinetic term or with more scalar fields\cite{Kaiser:2013sna}. It seems reasonable to suppose that in analogy with the case of the non-minimal coupling, the $\beta$-function formalism can be coherently applied to these models as well.

\section*{ Acknowledgements}
I would like to thank Nathalie Deruelle, David Kaiser, Andrei Linde and Diderik Roest for their suggestions. In particular I would like to thank Pierre Bin\'etruy for all the useful discussions that lead to the production of this work. I acknowledge the financial support of the UnivEarthS
Labex program at Sorbonne Paris Cit\'e (ANR-10-LABX-0023 and ANR-11-IDEX-0005-02) and the Paris Centre for Cosmological Physics.

\newpage
\appendix
\section*{Appendix}
\section{Some explicit examples.}
\label{Appendix}
In Sec.\ref{sec:general_case}, we discussed the consequences of introducing a further functional freedom in the Jordan frame forumalation of the model. In particular we considered:
\begin{equation}
  \label{non_minimal_action_3}
    S =\int\mathrm{d}^4x\sqrt{-g}\left( -  \frac{\Omega(\phi)}{2\kappa^2}R +  X - V_J(\phi) \right),
  \end{equation}
where $\Omega(\phi)$ and $V_J (\phi)$ have been defined as:
\begin{equation}
\Omega(\phi) = 1 + \xi f(\phi) ,\qquad \qquad \qquad V_J(\phi) = \lambda^2 g^2(\phi).
\end{equation}
After the usual conformal transformation we recover the Einstein frame formulation of the theory. By means of a field redefinition we are finally able to describe the system in terms of a field with a canonically normalized kinetic term. The strong coupling expression for $\tilde{f}(\varphi)$ is fixed by eq.\eqref{general_vaphi} and thus the model definition reduces to fixing an explicit expression for $g(\phi)$. In this appendix we consider some parametrizations for $\tilde{g}(\varphi)$ to study the possibility of preserving and evading the attractor. In particular we show that different universality classes can be reached.
\begin{itemize}
\item \textbf{Exponential.}
Let us consider:
\begin{equation}
 \tilde{g}(\varphi) =  \exp\left[ \sqrt{\frac{2}{3}} \left( \varphi - \varphi_{\mathrm{f}} \right) - \frac{e^{-\alpha (\varphi - \varphi_{\mathrm{f}})}}{\alpha}  \right].
\end{equation}
It is straightforward to derive the lowest order expression for $\beta(\varphi)$:
\begin{equation}
 \beta(\varphi) \sim - \frac{2 e^{-\alpha (\varphi - \varphi_{\mathrm{f}})}}{\alpha}. 
\end{equation}
This expression for $\beta(\varphi)$ corresponds to the exponential class presented in \cite{Binetruy:2014zya}. In this case the predictions for $n_s$ and $r$ are:
  \begin{eqnarray} 
            n_{s} &\simeq& 1 - \frac{2}{N}, \\
            r & \simeq& \frac{8}{\alpha^2 N^2}.
    \end{eqnarray}
It may be interesting to notice that the attractor at strong coupling of \cite{Kallosh:2013tua} can only be reproduced for $\alpha = \sqrt{2/3}$. For any other value for $\alpha$ the attractor is evaded.

\item \textbf{Chaotic.}
Let us consider:
\begin{equation}
 \tilde{g}(\varphi) = \left( \varphi - \varphi_{\mathrm{f}} \right)^\alpha \exp\left[ \sqrt{\frac{2}{3}} \left( \varphi - \varphi_{\mathrm{f}} \right) \right]
\end{equation}
clearly $ \tilde{g}_{,\varphi}(\varphi) / \tilde{g}(\varphi) = \sqrt{2/3} + \alpha /( \varphi - \varphi_{\mathrm{f}})$ that gives the lowest order expression:
\begin{equation}
\label{beta_general_ex}
\beta(\varphi) = \frac{- 2\alpha}{ \varphi - \varphi_{\mathrm{f}}}.
\end{equation}
This case corresponds to the Chaotic class discussed in \cite{Binetruy:2014zya} and gives:
  \begin{eqnarray} 
            n_{s} &\simeq &1 - \frac{2+a}{2N}, \\
            r & \simeq & \frac{4a}{N^2}.
    \end{eqnarray}
In this case the attractor at strong coupling is clearly evaded.

\item \textbf{Polynomial.}
Let us consider:
\begin{equation}
 \tilde{g}(\varphi) = \left\{  1 + \frac{P_1(\varphi)}{\exp \left[ \sqrt{\frac{2}{3}} \left( \varphi - \varphi_{\mathrm{f}} \right) \right]} \right\} \exp\left[ \sqrt{\frac{2}{3}} \left( \varphi - \varphi_{\mathrm{f}} \right) \right]
\end{equation}

where $P_1(\varphi)$ is a polynomial in $\varphi$. It is possible to prove that in this case the lowest order expression for $\beta(\varphi)$ reads:
\begin{equation}
\beta(\varphi) \sim \frac{P_2(\varphi)}{  \tilde{f}(\varphi)},
\end{equation}
where $P_2(\varphi)$ is a polynomial in $\varphi$. It is possible to show that at the lowest order the expressions for $n_s$ and $r$ are:
        \begin{eqnarray} 
            n_{s} - 1 &\simeq& - \frac{2}{N}, \\
            r & \simeq& \frac{12}{N^2}  .
    \end{eqnarray}
In this case the attractor is always preserved independently on the explicit expression for $P_1(\varphi)$. 
\end{itemize}


\addcontentsline{toc}{section}{References}
\begin{thebibliography}{99}

\def\hri#1#2{\href{http://arxiv.org/abs/#1}{[ArXiv:#1]#2}}
\def\hre#1#2{\href{http://arxiv.org/abs/#1/#2}{[ArXiv:#1/#2]}}
\def\hspi#1#2{\href{http://www.slac.stanford.edu/spires/find/hep/www?irn=#1}{#2}}

\bibitem{Ade:2015lrj} 
  P.~A.~R.~Ade \textit {et al.} [Planck Collaboration],
  arXiv:1502.02114 [astro-ph.CO].


\bibitem{Linde:1983gd} 
  A.~D.~Linde,
  Phys.\ Lett.\ B {\textbf{ 129}}, 177 (1983).



\bibitem{Bezrukov:2007ep} 
  F.~L.~Bezrukov and M.~Shaposhnikov,
  Phys.\ Lett.\ B {\textbf{ 659}}, 703 (2008)
  [arXiv:0710.3755 [hep-th]].


\bibitem{Starobinsky:1980te} 
  A.~A.~Starobinsky,
  Phys.\ Lett.\ B {\textbf{ 91}}, 99 (1980).


\bibitem{Roest:2013fha} 
  D.~Roest,
  JCAP {\textbf{ 1401}}, no. 01, 007 (2014)
  [arXiv:1309.1285 [hep-th]].

\bibitem{Mukhanov:2013tua} 
  V.~Mukhanov,
  Eur.\ Phys.\ J.\ C {\textbf{ 73}}, 2486 (2013)
  [arXiv:1303.3925 [astro-ph.CO]].


\bibitem{Binetruy:2014zya} 
  P.~Binetruy, E.~Kiritsis, J.~Mabillard, M.~Pieroni and C.~Rosset,
  JCAP {\textbf{ 1504}}, no. 04, 033 (2015)
  [arXiv:1407.0820 [astro-ph.CO]].

\bibitem{McFadden:2010na} 
  P.~McFadden and K.~Skenderis,
  J.\ Phys.\ Conf.\ Ser.\  {\textbf{ 222}}, 012007 (2010)
  [arXiv:1001.2007 [hep-th]].

\bibitem{Maldacena:1997re} 
  J.~M.~Maldacena,
  Int.\ J.\ Theor.\ Phys.\  {\textbf{ 38}}, 1113 (1999)
  [Adv.\ Theor.\ Math.\ Phys.\  {\textbf{ 2}}, 231 (1998)]
  [hep-th/9711200].


\bibitem{Kiritsis:2013gia} 
  E.~Kiritsis,
  JCAP \textbf{1311}, 011 (2013)
  [arXiv:1307.5873 [hep-th]].

\bibitem{Kallosh:2013tua} 
  R.~Kallosh, A.~Linde and D.~Roest,
  Phys.\ Rev.\ Lett.\  {\textbf{ 112}}, no. 1, 011303 (2014)
  [arXiv:1310.3950 [hep-th]].



\bibitem{Maldacena:2002vr} 
  J.~M.~Maldacena,
  JHEP {\textbf{ 0305}}, 013 (2003)
  [astro-ph/0210603].





\bibitem{Kallosh:2010ug} 
  R.~Kallosh and A.~Linde,
  JCAP {\textbf{ 1011}}, 011 (2010)
  [arXiv:1008.3375 [hep-th]].


\bibitem{Salopek:1990jq} 
  D.~S.~Salopek and J.~R.~Bond,
  Phys.\ Rev.\ D {\textbf{ 42}}, 3936 (1990).

\bibitem{Vennin:2014xta} 
  V.~Vennin,
  Phys.\ Rev.\ D {\textbf{89}}, no. 8, 083526 (2014)
  doi:10.1103/PhysRevD.89.083526
  [arXiv:1401.2926 [astro-ph.CO]].

\bibitem{Hoffman:2000ue} 
  M.~B.~Hoffman and M.~S.~Turner,
  Phys.\ Rev.\ D {\textbf{64}}, 023506 (2001)
  doi:10.1103/PhysRevD.64.023506
  [astro-ph/0006321].

\bibitem{Kinney:2002qn} 
  W.~H.~Kinney,
  Phys.\ Rev.\ D {\textbf{66}}, 083508 (2002)
  doi:10.1103/PhysRevD.66.083508
  [astro-ph/0206032].

\bibitem{Liddle:2003py} 
  A.~R.~Liddle,
  Phys.\ Rev.\ D {\textbf{68}}, 103504 (2003)
  doi:10.1103/PhysRevD.68.103504
  [astro-ph/0307286].

\bibitem{Kallosh:2013hoa} 
  R.~Kallosh and A.~Linde,
  JCAP \textbf{ 1307}, 002 (2013)
  doi:10.1088/1475-7516/2013/07/002
  [arXiv:1306.5220 [hep-th]].

\bibitem{Kallosh:2013daa} 
  R.~Kallosh and A.~Linde,
  JCAP \textbf{ 1312}, 006 (2013)
  doi:10.1088/1475-7516/2013/12/006
  [arXiv:1309.2015 [hep-th]].

\bibitem{Kallosh:2013yoa} 
  R.~Kallosh, A.~Linde and D.~Roest,
  JHEP \textbf{ 1311}, 198 (2013)
  doi:10.1007/JHEP11(2013)198
  [arXiv:1311.0472 [hep-th]].

\bibitem{Kallosh:2015lwa} 
  R.~Kallosh and A.~Linde,
  Phys.\ Rev.\ D \textbf{ 91}, 083528 (2015)
  doi:10.1103/PhysRevD.91.083528
  [arXiv:1502.07733 [astro-ph.CO]].


\bibitem{Galante:2014ifa} 
  M.~Galante, R.~Kallosh, A.~Linde and D.~Roest,
  Phys.\ Rev.\ Lett.\  \textbf{114}, no. 14, 141302 (2015)
  [arXiv:1412.3797 [hep-th]].


\bibitem{Kallosh:2014laa} 
  R.~Kallosh, A.~Linde and D.~Roest,
  JHEP \textbf {1409}, 062 (2014)
  [arXiv:1407.4471 [hep-th]].


\bibitem{Kaiser:2013sna} 
  D.~I.~Kaiser and E.~I.~Sfakianakis,
  Phys.\ Rev.\ Lett.\  \textbf{112}, no. 1, 011302 (2014)
  [arXiv:1304.0363 [astro-ph.CO]].

\end{thebibliography}
\end{document}